\documentclass{emulateapj}
\usepackage{epsfig}

\begin{document}

\slugcomment{\apj\ Letters, submitted 7 June 2011, accepted 1 August 2011 (Printed \today)}
\title{The Fate of Stellar Mass Loss in Central Cluster Galaxies}
\author{G. Mark Voit\altaffilmark{1} and Megan Donahue\altaffilmark{1}
         } 
\altaffiltext{1}{Department of Physics and Astronomy,
                 Michigan State University,
                 East Lansing, MI 48824, 
                 voit@pa.msu.edu}

\begin{abstract}
Star formation within the central galaxies of galaxy clusters is often interpreted as being fueled by cooling of the hot intracluster medium.  However, the star-forming gas is dusty, and {\em Spitzer} spectra show that the dust properties are similar to those in more normal star-forming environments, in which the dust has come from the winds of dying stars.  Here we consider whether the primary source of the star-forming gas in central cluster galaxies could be normal stellar mass loss.  We show that the overall stellar mass-loss rate in a large central galaxy ($\sim 4$--8$\, M_\odot \, {\rm yr}^{-1}$) is at least as large as the observed star-formation rates in all but the most extreme cases and must be included in any assessment of the gas-mass budget of a central cluster galaxy.  We also present arguments suggesting that the gas shed by stars in galaxy clusters with high core pressures and short central cooling times may remain cool and distinct from its hot surroundings, thereby preserving the dust within it.
\end{abstract}


\section{Introduction}

\setcounter{footnote}{0}

The purpose of this paper is to draw attention to an obvious but sometimes overlooked fact:  Normal stellar mass loss in the central galaxies of galaxy clusters releases substantial amounts of gas into their cores.  Our understanding of cluster cores cannot be considered complete unless it accounts for what happens to that gas.  Most of the galaxies at the centers of clusters are not forming new stars and lack large reservoirs of cold gas, implying that the matter shed by dying stars is usually expelled.  However, a significant minority of central cluster galaxies contain $10^9$--$10^{11} \, M_\odot$ of cold gas \citep{Edge01} that is forming stars at rates from $1 \, M_\odot \, {\rm yr}^{-1}$ to $10 \, M_\odot \, {\rm yr}^{-1}$, and in a few cases up to $100 \, M_\odot {\rm yr}^{-1}$ or more \citep[e.g.,][]{Odea+08}.  The literature on the origin of this star-forming gas has focused primarily on cooling and condensation from the hot phase as its main source, positing that gas flows from the intracluster medium into the central galaxy.  Here we argue that normal stellar mass loss is potentially a more important source of star-forming gas and that the usual direction of gas flow may be outward from the central galaxy, even in many clusters with short central cooling times.

Cooling and condensation from the hot phase has received far more attention than stellar mass loss because gas from red-giant winds and planetary nebulae in elliptical galaxies has generally been assumed to assimilate into the hot phase soon after being ejected.  The hydrodynamics of assimilation is complex.  Attempts to sketch out how it happens have postulated that Rayleigh-Taylor and Kelvin-Helmholtz instabilities rip the ejected gas clouds into small fragments that then evaporate through thermal conduction \citep{Mathews90}.  Exactly how this process of assimilation occurs in elliptical galaxies has not received much attention from simulators, and the simulations done to date have not provided conclusive answers about whether the ejected gas is quickly assimilated or remains cool and distinct from the hot ambient medium \citep{pb08,bp09}.

Recent observations are motivating us to reconsider the assumption that stellar ejecta are always rapidly heated and assimilated in elliptical galaxies.  For example, infrared spectra of star-forming central cluster galaxies observed with {\em Spitzer} exhibit PAH emission and far-IR characteristics similar to those of more normal star-forming galaxies \citep{Donahue+11}.  Dust has long been known to be present in the associated emission-line nebulae \citep[e.g.,][]{Sparks+89,dv93} and its presence is hard to explain if the only source of cool gas is condensation out of a hot medium in which the dust sputtering time is $\sim 1$~Myr \citep[but see][]{fjd94}. The presence of fragile PAHs is even harder to understand.  Furthermore, the relative strengths of the PAH features resemble those of star-forming regions in late-type galaxies, strongly suggesting that the dusty gas has a similar origin in the winds of dying stars and has remained below the $\sim 10^6$~K temperatures at which sputtering becomes efficient, because sputtering would eliminate the smallest grains first.

The ability of ejected stellar gas to remain cool, without assimilating into the hot ambient medium of an elliptical galaxy, is likely to depend on the pressure of that medium.  Higher ambient pressures, such as those found in the central galaxies of cool-core clusters, promote more rapid radiative cooling which will more effectively counteract heating of gas that is already cool.  The presence of cool gas in some of those galaxies might therefore be related more directly to the preservation of stellar ejecta in a cool state than to the ability of the hot gas to condense.

Our brief discussion of these issues in this {\em Letter} proceeds as follows.  Section~2 evaluates the overall gas budget in central cluster galaxies, showing that in most cases the total amount of normal stellar mass loss is comparable to or greater than the star-formation rate.  Section~3 examines whether gas ejected by stars into a hot elliptical galaxy can plausibly remain cold.  Section~4 concludes with a brief summary of the implications.  All observed quantities are for a flat $\Lambda$CDM cosmology with $H_0 = 70 \, {\rm km \, s^{-1} \, Mpc}$ and $\Omega_{\rm M} = 0.3$. 

\newpage

\section{Balancing the Gas Budget}
\label{sec-mbudget}

Star formation in a central cluster galaxy can tap two sources of gas, the hot intracluster medium and the stars of the galaxy itself.  The rate at which gas condenses from the hot medium and settles into the central galaxy is hard to measure and is certainly not as large as was once thought \citep{pf06}.  Observational limits on emission lines from intermediate-temperature gas are the primary constraints on that condensation rate and yield inconclusive results, since the upper limits on mass-cooling rates tend to decline as one looks at emission lines characteristic of progressively cooler temperatures \citep[e.g.,][]{Sanders+10}.  It has also been suggested that turbulent mixing can transfer gas from the hot phase to the cold phase without producing emission lines from intermediate temperatures \citep[e.g.,][]{Fabian+02}.  However, the total amount of intracluster gas that has condensed in a central cluster galaxy during the last few billion years cannot exceed the mass in cold gas, plus the mass in young to intermediate-age stars, minus the mass released by the older stellar population.

The total mass-loss rate from the older population is not directly measurable but can be estimated from the properties of the central galaxy.  Assuming an initial mass function (IMF) and a population age, one can calculate the mass of stars that have died during a given time interval and subtract the mass of their remnants.  \citet{lk11} have compiled results for different IMFs that can be represented with a simple fitting formula introduced by \citet{Jungwiert+01}, in which the fraction of the initial stellar mass that has been returned to interstellar space by time $t$ equals $f_g = C_0 \ln ( t\lambda^{-1} + 1)$, with $C_0 \approx 0.05$ and $\lambda \approx 5$~Myr, depending on the IMF.  The specific mass-loss rate per initial stellar mass is then $\dot{f}_g = C_0 / (t + \lambda)  \approx 4 \times 10^{-12} \, {\rm yr}^{-1}$ for a population that is $\sim 12$~Gyr old.  If there are significant numbers of younger stars, the specific mass-loss rate will be greater than this.

Assessments of the total stellar mass currently associated with a central cluster galaxy depend somewhat on where one draws the boundary between the central galaxy and the intracluster light.  Here we will base our estimates of the total stellar mass in central cluster galaxies on the $I$-band observations of \citet{Gonzales+05}, who find average values of $M_I (<10 \, {\rm kpc}) = -23.11$, $M_I (<50 \, {\rm kpc}) = -24.26$, and $M_I (<300 \, {\rm kpc}) = -24.96$.  According to the population synthesis models of \citet{bc03}, a stellar population with an initial mass of $10^{12} \, M_\odot$, a \citet{Chabrier03} IMF, and an age of 12~Gyr would now have $M_I = -24.15$.  The current stellar population within each radius therefore corresponds to an {\em initial} stellar mass of $M_{\rm init}(<10 \, {\rm kpc}) = 4 \times 10^{11} \, M_\odot$, $M_{\rm init}(<50 \, {\rm kpc}) = 1.1 \times 10^{12} \, M_\odot$, and $M_{\rm init}(<300 \, {\rm kpc}) = 2.1 \times 10^{12} \, M_\odot$.

\begin{figure}[t]
\includegraphics[width=3.35in, trim = 0in 0in 0in 0in]{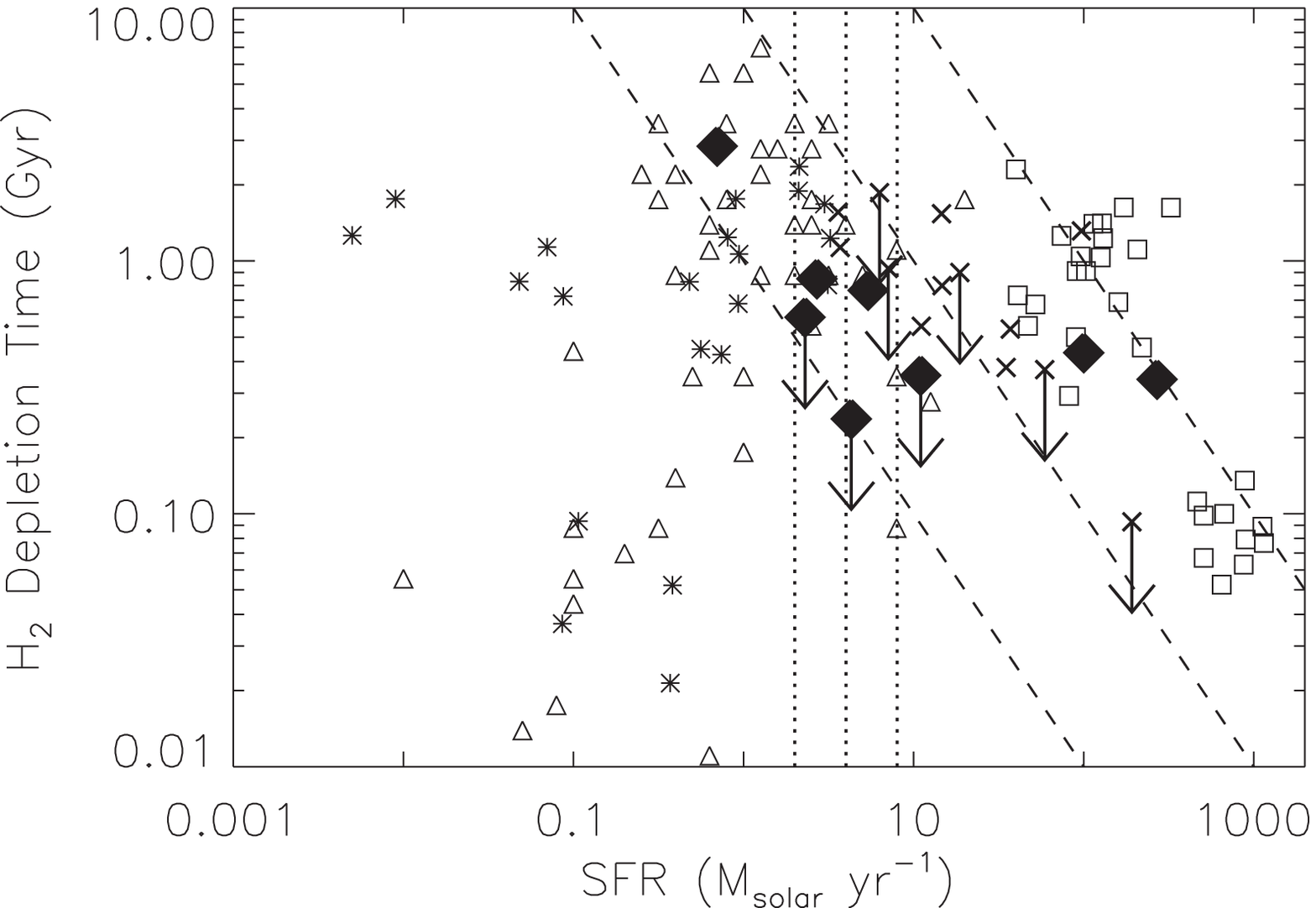} \\
\caption{ \footnotesize 
Star formation rates and gas depletion times in central cluster galaxies and other star-forming galaxies.  
The star-formation rates for central cluster galaxies are from \citet{Donahue+11} (diamonds) and \citet{Odea+08} (crosses), with gas masses from \citet{Edge01} used to obtain depletion times.  Asterisks and triangles represent nearby star-forming galaxies from \citet{Leroy+08} and \citet{Krumholz+11}, respectively.  Squares represent star-forming galaxies at $z>1$ from \citet{Genzel+10}.  Dotted lines show rates of 2, 4, and $8 \,M_\odot \, {\rm yr}^{-1}$, respectively.  Dashed lines show H$_2$ gas masses of $10^9$, $10^{10}$, and $10^{11} \, M_\odot$ (left to right). 
\vspace*{1em}
\label{fig-tsfr}}
\end{figure}

Multiplying these numbers by the specific mass-loss rate given above implies current mass-loss rates within central cluster galaxies $\sim 2 \, M_\odot \, {\rm yr}^{-1}$ at $< 10$~kpc,  $\sim 4 \, M_\odot \, {\rm yr}^{-1}$ at $< 50$~kpc, and $\sim 8 \, M_\odot \, {\rm yr}^{-1}$ at $< 300$~kpc.  All of these numbers increase if there are stars younger than 12~Gyr.  Figure~\ref{fig-tsfr} shows how this source of mass compares with the star-formation rates inferred from {\em Spitzer} observations by \citet{Donahue+11} and \citet{Odea+08}.  These samples suffer from a strong selection bias, in that they were selected for {\em Spitzer} followup because they were expected to be particularly bright infrared sources.  Nevertheless, only the most extreme examples have star-formation rates that greatly exceed their stellar mass-loss rates.  Most central cluster galaxies have no measurable star formation, and systems with more than $\sim 10 \, M_\odot \, {\rm yr}^{-1}$ of star formation are rare.  In other words, the observed star-formation rates in central cluster galaxies are usually similar to or less than the total stellar mass-loss rates, with some notable exceptions.  Stellar mass-loss can account for these exceptional cases only if they are brief bursts of rapid star formation in systems with a lower long-term average rate.  Alternatively, they may be cases in which ICM condensation is the dominant mass source, in which case one might expect them to have a lower dust-to-gas ratio---a hypothesis that can potentially be tested with {\em Herschel} observations. 

The vertical axis of Figure~\ref{fig-tsfr} shows estimates of the depletion time for cold star-forming gas inferred from the quotient of the molecular gas mass in the central cluster galaxy, inferred from CO observations by \citet{Edge01}, and the galaxy's star-formation rate.  Gas depletion times for star-forming central cluster galaxies are typically $\sim 1$~Gyr, interestingly similar to those of other star-forming galaxies in very different environments but with comparable star-formation rates.   Asterisks and triangles in Figure~\ref{fig-tsfr} represent nearby star-forming galaxies from \citet{Leroy+08} and \citet{Krumholz+11}, respectively, while the squares represent star-forming galaxies at $z>1$ from \citet{Genzel+10}.  The fact that central cluster galaxies occupy a similar region in the star-formation/gas depletion plane suggests that similar mechanisms may be regulating star formation in all of these systems. 

\newpage

\section{Thermal History of Stellar Ejecta}
\label{sec-thermhist}

The estimates made in the previous section do not prove that normal stellar mass-loss is a more important source of star-forming gas than condensation from the hot medium.  They do, however, indicate that it cannot be neglected.   Here we discuss what happens to that gas after it is ejected and how its fate may depend on the pressure of the hot ambient medium. 

First, consider what happens in clusters with a central electron density $n_{e0} \lesssim 0.02 \, {\rm cm}^{-3}$, implying a central cooling time $\gtrsim 2$~Gyr.  The total stellar mass in the inner $\sim 20$~kpc of a central cluster galaxy is nearly independent of a cluster's global properties \citep[e.g.,][]{Haarsma+10}, meaning that central stellar mass loss rates do not differ much from cluster to cluster.  Over the last $\sim 6$~Gyr, a central galaxy's stars have released $\sim 3 \times 10^{10} \, M_\odot$ within a $\sim 50$~kpc radius and $\sim 10^{10} \, M_\odot$ within $\sim 10$~kpc.  However, clusters with central cooling times $\gtrsim 2$~Gyr do not have have star formation rates $> 1 \, M_\odot \, {\rm yr}^{-1}$ and do not have cold-gas reservoirs with $> 10^{10} \, M_\odot$.  Lacking another mass sink, we conclude that the ejected gas in these systems must have been heated to the ambient temperature of the intracluster medium, in accord with previous expectations \citep{Mathews90}.  Most of it has subsequently been transported away from the cluster's center, since the gas mass within 10~kpc is $\lesssim 10^{9} \, M_\odot$.  However, this class of clusters tends not to show much radio power from a central AGN \citep{Cavagnolo+08}, suggesting that a mechanism other than AGN feedback, perhaps a combination of ICM turbulence and thermal conduction, is responsible for the transport. 

Next, consider clusters with greater central electron densities and shorter central cooling times, but with star-formation rates $\lesssim 1 \, M_\odot \, {\rm yr}^{-1}$ and cold-gas reservoirs $< 10^{10} \, M_\odot$.  Again, the primary sink for normal stellar mass loss would appear to be the hot ambient medium.  Furthermore, gas-mass accounting suggests that the net flow of gas is still outward, at least for systems with $n_{e0} \lesssim 0.1 \, {\rm cm}^{-3}$, because the mass of hot gas within 10~kpc is $\lesssim 10^{10} \, M_\odot$.   It is also possible that some of the ejected stellar mass has accreted onto a central black hole.   In either case, the gas flow pattern in the central 10~kpc is not simply an AGN-regulated cooling flow that gradually transports the hot ambient medium inward.  If condensation of the intracluster medium is fueling the AGN, then the resulting feedback must be carrying most of the ejected stellar mass {\em outward} from the central 10~kpc, at least in systems with $n_{e0} \lesssim 0.1 \, {\rm cm}^{-3}$ and no other comparable mass sinks.

Now we turn our attention to clusters with $n_{e0} \lesssim 0.1 \, {\rm cm}^{-3}$ and star-formation rates $\gtrsim 1 \, M_\odot \, {\rm yr}^{-1}$, in which the fate of stellar mass loss is potentially more interesting.  The main question to ask is whether mass flowing from stars in the central galaxies of these clusters can remain cool and distinct from the hot ambient medium long enough to collect into a cool-gas reservoir that fuels star formation.  This {\em Letter} will not definitively answer that question but will present arguments suggesting that preservation is possible and depends on the pressure of the ambient medium.

Most of the matter returning to interstellar space is coming from evolved stars at speeds not greatly exceeding the star's escape velocity and will remain cool during its initial interaction with the ambient medium.  Here we will assume that a typical dying star is losing $\dot{m} = (10^{-7} \, M_\odot \, {\rm yr}^{-1}) \dot{m}_{-7}$ with a wind speed of $v_{\rm w} = (40 \, {\rm km \, s^{-1}})  v_{40} $.  Suppose the star is moving at $v_* = (400 \, {\rm km \, s^{-1}})  v_{400} $ relative to an ambient medium with electron density $(0.1 \, {\rm cm}^{-3}) n_{0.1}$.  The star's motion may or may not be supersonic, depending on whether the ambient temperature is greater than $(0.6 \, {\rm keV})v_{400}^2$. In most clusters of interest the central temperature is hotter than this, the star's motion will be subsonic, and the stellar wind will shock when its ram pressure approaches the ambient pressure, at a radius $r_s \approx (7.5 \times 10^{16} \, {\rm cm}) \dot{m}_{-7}^{1/2} v_{40}^{1/2} n_{0.1}^{-1/2} T_{\rm keV}^{-1/2}$, where $T_{\rm keV}$ is the ambient temperature in units of keV.  After passing through this shock at the stellar-wind speed, the temperature of the stellar ejecta will not be much greater than $10^4$~K, but what happens next?

The standard assumption is that Rayleigh-Taylor and Kelvin-Helmholz instabilities will break the dense ejected gas into fragments that quickly mix with the hot ambient medium \citep[e.g.,][]{Mathews90}.  These instabilities do indeed develop in hydrodynamical simulations of stellar wind interactions with a hot medium, which also show that that radiative cooling can act to delay mixing \citep{pb08,bp09}.  However, such treatments assume that the fluid approximation can be applied to this problem.  In fact, the ion mean free path in the ambient plasma of a cluster's core is at least two orders of magnitude larger than the standoff radius of the stellar wind, implying that the interaction between the wind and the ambient medium must be magnetically mediated.  In that case, the usual hydrodynamical instabilities may be suppressed by magnetic draping \citep[e.g.,][]{Dursi07}, which can also insulate the stellar ejecta from thermal conduction.  One possibly relevant example is the evolved star Mira, whose wind enters a bow shock $\sim 1.6 \times 10^{17} \, {\rm cm}$ (0.05~pc) from the star and then flows into a trailing tail.  Far-UV emission from the tail, which extends at least 4~pc behind the star, is presumed to be from molecular hydrogen, indicating that at least some of the stellar ejecta remains cold and dense \citep{Martin+07}.  Ultimately, the survival of ejected gas in a cold state depends on its magnetic connectivity with the ambient medium, which has not yet been adequately simulated, but in general, higher ambient pressure will lead to greater gas density in the tail, making it easier for radiative cooling to offset thermal conduction.

Another consideration is whether high-speed collisions with the winds of other mass-losing stars can effectively heat cold ejected gas.  One way to assess the importance of these events is to estimate the fractional volume swept out by the cross sections of stellar-wind standoff shocks.  The standoff shock of each star's wind has a cross-sectional area $\sigma_w = 0.25 \, \dot{m} v_w P_a^{-1}$, where $P_a = 3 \times 10^{-10} \, n_{0.1} T_{\rm keV} \, {\rm erg \, cm^{-3}}$ is the pressure of the ambient medium.  
Assuming a stellar mass-loss rate $\sim 2 \, M_\odot \, {\rm yr}^{-1}$ within 10~kpc of the galaxy's center, we find that the fractional volume visited by stellar-wind shocks in a Hubble time is $\sim 0.1 v_{400} v_{40} n_{0.1}^{-1} T_{\rm keV}^{-1}$.  This is not a rigorous calculation but rather is meant to illustrate that ambient pressure again plays an important role.  Stellar winds in clusters with low central entropy and high central pressure sweep out only a small fraction of the volume within the central 10~kpc,  while winds from stars in galaxies with ambient pressures $\lesssim 10^{-2} \, {\rm keV \, cm^{-3}}$ can sweep out a large fraction of that volume, potentially preventing cool stellar ejecta from accumulating and forming stars.  A more detailed quantitative assessment of this mechanism is needed to determine how the survival of cold clouds depends on ambient pressure and the local density of mass-losing stars. 

\section{Discussion}
\label{sec-discussion}

We have shown that the total stellar mass-loss rate in a central cluster galaxy is comparable or greater than the star-formation and inferred gas-cooling rates in all but the most extreme cluster cores.  We have also presented evidence that gas ejected from stars in cluster cores might remain cold, particularly in clusters with low core entropy and high core pressure.  Taken together, these findings suggest that the main source of the cool star-forming gas found in many low-entropy cluster cores may be the stars of the central galaxy, providing a natural explanation for why that gas is dusty.  Even if some of the star-forming gas has condensed from the intracluster medium, the contribution of stellar mass loss cannot be neglected.  It has long been an important source term in models of cooling flows in individual elliptical galaxies \citep[see][and references therein]{mb03}.  However, the larger stellar masses and greater ambient pressures of central cluster galaxies may make them much more effective at recycling stellar ejecta into new stars.  Future simulations of central cluster galaxies and their relationship to cool cluster cores must account for this important source of cool gas.

\vspace*{1.0em}

The authors thank Andrey Kravtsov and the KITP cool-core discussion group for their input.  This work was partially supported at MSU by NSF grant AST-0908819.  It was initiated at the Kavli Institute for Theoretical Physics in Santa Barbara, supported in part by the National Science Foundation under Grant No. NSF PHY05-51164.
 

\end{document}